\documentstyle[epsfig,12pt,preprint,tighten,aps]{revtex}
\begin{document}

\draft

\title{\rightline{{\tt August 2000}}
\ \\
Maximal $\nu_\mu - \nu_\tau$ oscillations, 
the see-saw mechanism and the Exact Parity Model 
}
\author{T. L. Yoon\footnote{tlyoon@physics.unimelb.edu.au}
 and R. Foot\footnote{foot@physics.unimelb.edu.au}}
\address{
School of Physics\\
Research Centre for High Energy Physics\\
The University of Melbourne\\
Victoria 3010 Australia }

\maketitle

\begin{abstract}
We examine one simple mechanism which leads to approximate maximal 
$\nu_\mu - \nu_\tau$ oscillations  
in the standard see-saw model. In particular, we show that 
this scheme could be implemented in the Exact Parity Model
(also known as the mirror matter model).
Within this framework 
the solar neutrino problem is solved by maximal
$\nu_e \to \nu'_e$ oscillations (with $\nu'_e$ is the essentially
sterile mirror partner of $\nu_e$) and the LSND
evidence for $\nu_e \to \nu_\mu$ oscillations
can also be explained.

\end{abstract}

\newpage
\vskip 0.5cm
Neutrino physics continues to provide the most promising window
on physics beyond the standard model. The
evidence comes from three different 
classes of experiments, namely the atmospheric\cite{atmos,sk}, 
solar\cite{solar} and 
LSND\cite{lsnd} experiments. Of these, the most convincing evidence 
of neutrino oscillation comes from the atmospheric neutrino anomaly 
as confirmed by SuperKamiokande\cite{sk}.
At the present time the only two flavour oscillation solutions
consistent with the superKamiokande data are
$\nu_\mu \to \nu_\tau$ and $\nu_\mu \to
\nu_{sterile}$ oscillations\cite{fvy}.
Recently the superKamiokande collaboration 
have argued that the $\nu_\mu \to \nu_{sterile}$ solution is 
disfavoured at more than $99\%$ C.L\cite{sk2}. However 
this conclusion depends on how the data is analysised 
and should not be considered as conclusive\cite{foot00}.
Furthermore a global analysis of the 
$\nu_\mu \to \nu_{sterile}$ solution to the atmospheric
neutrino data provides a reasonable fit\cite{concha}.
Thus, the $\nu_\mu \to \nu_{sterile}$ oscillations remain a viable
explanation of the atmospheric neutrino data.
However, for the purposes of this paper
we will assume that nature
chooses the $\nu_\mu \to \nu_\tau$ oscillation solution.
The main point of our paper is to point out a rather simple
way of achieving approximately maximal $\nu_\mu \to \nu_\tau$
oscillations in the context of the usual see-saw model in
a natural way.  
We also point out how the
solar neutrino and LSND indications for neutrino oscillations
could also be understood in this scheme.

In the context of the standard see-saw model,
it is well known that the effective mass matrix of 
the light left-handed neutrinos ${\bf m}_L$ can be 
obtained from diagonalising the seesaw mass matrix 
\begin{equation} 
{\bf M} = \left(\begin{array}{cc}
0&\ {\bf M}_D\\
\ {\bf M}_D^T&{\bf M}_R
\end{array}\right),
\label{1}
\end{equation}
leading to
\begin{equation}
{\bf m}_L = {-V_L d_\nu W_R D^{-1} W_R^{T} d_\nu V_L^{T}}.
\label{2}
\end{equation}
In Eq. (\ref{1}) and Eq. (\ref{2}), ${\bf M}_D$ 
and ${\bf M}_R$ are the ${3 \times 3}$ Dirac and Majorana matrix for 
${\nu_L}$ and ${\nu_R}$ respectively. $W_R = {V_R}^\dagger {U_R}^*$
where $V_{R,L}$ and $U_R$ are unitary matrices that diagonalise
${\bf M_D}$ and ${\bf M_R}$ as 
\begin{equation}
{\bf M}_D = V_L d_\nu V_R^{\dagger}, \
{\bf M}_R = U_R D U_R^T,
\label{3}
\end{equation}
where $d_\nu$ and $D$ are diagonal matrices 
containing real eigenvalues of ${\bf M}_D$ and ${\bf M}_R $, i.e. 
\begin{equation}
d_\nu = diag(m^D_1, m^D_2 ,m^D_3), \;
D = diag(M_1,M_2,M_3). 
\label{4}
\end{equation}
From general considerations, it seems most natural 
to assume that ${\bf M}_D$ is hierarchical.\footnote{Although 
it is not a necessary condition in the scheme we shall propose here.} 
In addition, 
it is also most natural to expect that ${\bf M}_D$ 
is approximately aligned with 
the charged lepton mass matrix, 
as a parallel to the charged quark sector. In fact, these
features arise in simple GUT models such as $SO(10)$\cite{georgi} as well
as in models with quark lepton symmetry\cite{ps,fl}. 
For ${\bf M}_R$, however, there 
is no natural expectation as such because the charged leptons 
and quarks have 
no Majorana mass due to electric charge conservation. In contrast to 
the Dirac mass terms in the Standard Model, 
${\bf M}_R$ is expected to arise from quite different physics,
e.g. from coupling with a distinct 
Higgs field at a much higher scale. We can then ask
the question: Is there any reasonable choice for ${\bf M}_R$
that could lead to approximately maximal 
$\nu_{\mu L} - \nu_{\tau L}$ oscillations compatible with the
atmospheric neutrino anomaly? 

Let's consider the 2-3 sector of the mass matrices. 
The symmetric Majorana mass matrix of $\nu_R$ can be generally 
written as\footnote{
We use the basis with $V_R = 1$, which we can choose without loss
of generality.}
\begin{equation}
{\bf M}_R = \left(
\begin{array}{cc}
\lambda_1&\ \lambda_3\\
\ {\lambda_3}&{\lambda_2}
\end{array}\right),
\label {5}
\end{equation}
and is diagonalised by 
\begin{equation} 
U_R = \left(\begin{array}{cc}
\cos \, \psi &\ -\sin \, \psi \\
\ \sin \, \psi & \cos \, \psi
\end{array}\right).
\label{6}
\end{equation}
The mixing angle among $\nu_{{\mu}R} - \nu_{{\tau}R}$ is parametried by 
\begin{equation}
\tan \, 2\psi = \frac{2\lambda_3}{\lambda_1 - \lambda_2}, 
\label{7}
\end{equation}
and the eigenvalues are given by
\begin{equation}
M_{2,3} = \frac{1}{2} \,
( \lambda_1 + \lambda_2 ) \mp \frac{1}{2} \sqrt{( \lambda_2  - \lambda_1 )^2
+ 4 \, \lambda_3^2 } \,.
\label{8}
\end{equation}
The mixing angle among $ \nu_{{\mu}L} - \nu_{{\tau}L} $
implied by Eq. (\ref{2}) is easily worked out to be 
\begin{equation}
\tan 2\theta_L = 
\frac{2m^D_2 m^D_3 \sin \psi \cos \psi (\frac{1}{M_3} - \frac{1}{M_2})}
{ ({m^D_3})^2(\frac{\sin^2 \psi}{M_2} + \frac{\cos^2 \psi}{M_3}) - 
  ({m^D_2})^2(\frac{\cos^2 \psi}{M_2} + \frac{\sin^2 \psi}{M_3}) }. 
\label{9}
\end{equation}
Approximate maximal mixing requires
the condition
\begin{equation}
\tan^2 2\theta_L \gg 1.
\label{10}
\end{equation}
As already mentioned, we will assume that
${\bf M}_D$ is diagonal (i.e $V_L = 1$),
(needless to say, our conclusion requires only that
$V_L$ is approximately diagonal as in the case of the CKM matrix).
The issue
is to find the form of ${\bf M}_R$ which then leads to maximal oscillations
for the light $\nu_{\mu L} \to \nu_{\tau L}$ states.
One possibility already discussed in the literature\cite{Akh} is that
\begin{equation}
{\bf M}_R^{-1} \propto 
\left( \begin{array}{cc}
1&\ p\\
\ p&p^2
\end{array} \right),
\label{11}
\end{equation}
where $p = \frac {m_2^D}{m_3^D}$. This case has been studied in Ref.\cite{Akh}
in detail. 
In this case the light neutrinos are hierarchical.
While this is certainly an interesting possibility, it doesn't seem to be
the simplest possibility in our opinion. In particular,
as far as we can see there is no obvious approximate symmetry
corresponding to the form Eq.(\ref{11}) (although the authors
of Ref.\cite{Akh} argue that it is more natural than it looks).

It seems to us that the most natural choice of ${\bf M}_R$ which does the
trick is where 
${\bf M}_R$ is  approximately maximally mixed, with 
\begin{equation}
|\lambda_3| \gg |\lambda_2|, |\lambda_1|.
\label{140}
\end{equation}
In this limit, the heavy Majorana mass eigenvalues are
[from Eq. (\ref{8})],
\begin{equation}
M_{2,3} \simeq \mp \lambda_3 + \frac{\lambda_1+\lambda_2}{2},
\label{150}
\end{equation}
which are approximately degenerate.  It is easy to show that
${\bf M}_R$ is diagonalised by 
the unitary transformation, Eq.(\ref{6}) with
\begin{equation}
\tan^2 \psi  = 1 - {(\lambda_1 - \lambda_2)\over \lambda_3}
+ {\cal O}\left({\lambda_{1,2}\over \lambda_3}\right)^2.
\end{equation}
It is easily checked that indeed, in the limit of Eq. (\ref{140}),
\begin{equation}
\tan^2 2\theta_L \simeq \left({-2\lambda_3 m_2^D m_3^D \over
(m_3^D)^2 \lambda_2 - (m_2^D)^2 \lambda_1 } \right)^2 \gg 1, 
\end{equation}
irrespective of whether there is hierarchy in $m_{2,3}^D$.
Hence maximal mixing for the light $\nu_{\mu,\tau}$ states
follows in the limit $|\lambda_3| \gg |\lambda_1|, |\lambda_2|$. 
Meanwhile, the effective $\nu_L$ light Majorana mass matrix has
the form:
\begin{equation}
{\bf m}_L \sim \
\left( \begin{array}{cc} (m_2^{D})^2\left({-\lambda_1 \over
\lambda_3^2}\right)
\null & \null 
\frac{m^D_2 m^D_3}{\lambda_3} \\
\frac{m^D_2 m^D_3}{\lambda_3} \null & \null 
(m_3^{D})^2\left({-\lambda_2 \over
\lambda_3^2}\right)
\end{array} \right),
\label{175}
\end{equation}
resulting in an approximate degenerate pair of light flavour neutrino masses 
(where we have made the usual phase transformation to make
them positive) of
\begin{equation}
m_L \simeq \frac{m^D_2 m^D_3}{\lambda_3} \pm {1 \over 2}
\left( (m_2^D)^2{\lambda_1 \over \lambda_3^2} + 
(m_3^D)^2{\lambda_2 \over \lambda_3^2}\right).
\label{178}
\end{equation}
Observe that in this 
scenario there is an approximate $U(1)_{L_\mu - L_\tau}$
global symmetry, which may simply be an accidental approximate 
symmetry of the theory. Nevertheless this approximate global
symmetry could play an important role in preventing radiative
corrections from spoiling the form of ${\bf m}_L$. 

Of course, 
this see-saw model with an approximately
diagonal ${\bf M}_D$ but with the ``lop sided'' ${\bf M}_R$ 
is not the only way of achieving approximately
maximal $\nu_\mu \to \nu_\tau$ mixing.
However, it does seem rather nice to us.  As mentioned already,
this scheme could easily
be implemented in standard GUT models as well as models with
quark-lepton symmetry. Moreover, as we will show,
it fits in nicely with models with unbroken
parity symmetry (Exact Parity model).
While we arrived at this scheme independently, 
we have searched the literature and have discovered that
similar ideas are contained in Ref.\cite{allanach}
in the context of models with $U(1)_F$ family gauge symmetry. 
Let us now turn to the other neutrino anomalies: The solar and the
LSND experiments.

The LSND experiment suggests $\nu_e \to \nu_\mu$ oscillations 
with $\sin^2 \theta_{\mu e} \sim 10^{-2}$ and
$\delta m^2 \sim 1 \ eV^2$\cite{lsnd}.
This can easily be incorporated into this scheme. 
We need only assume that ${\bf M}_R$ has the approximate form:
\begin{equation} 
{\bf M}_R \sim \left(\begin{array}{ccc}
\alpha&0&0\\
0&0&\beta\\
0&\beta&0
\end{array}\right)
\end{equation}
where the zero entries are in general non-zero but small
(relatively to the other entries).

Within the context of our scheme, the solar neutrino anomaly
may be solved by oscillations into a light sterile
neutrino (and elegance suggests three such species, which
we denote by $\nu'_{e,\mu,\tau}$).
There are essentially two possibilities: Either small angle
$\nu_e \to \nu'_{e}$ 
MSW solution is invoked\cite{barger} or approximately 
maximal $\nu_e \to \nu'_{e}$ oscillations may
be responsible\cite{fv95}.
The latter can explain (and in fact predicted\cite{flv,fv95}) the 
approximate $50\%$ reduction
of solar neutrinos including the observed energy
independent superKamiokande flux suppression\cite{sup} (although
Homestake is a little on the low side).
Futhermore, observations from  Borexino and Kamland must
find at least one ``smoking gun'' signature if this is the 
physics responsible for
the solar neutrino deficit (and SNO should see no anomolous Neutral
current/charged current ratio)\cite{fv95}.
On the theoretical front,
an elegant motivation for maximal $\nu_e \to \nu'_e$
oscillations comes from
the observation that gauge models with unbroken parity symmetry
predict that each of the three ordinary neutrinos are approximately
maximally mixed with a sterile partner if neutrinos
have mass\cite{flv,flv2}.  
Obviously 
this would suggest that the most natural explanation for the 
atmospheric
neutrino anomaly is with maximal $\nu_\mu \to \nu'_{\mu}$
oscillations. Nevertheless it is still possible that 
the atmospheric neutrino anomaly may be predominately due to 
$\nu_\mu \to \nu_\tau$ oscillations
if the the oscillation length for $\nu_\mu \to \nu'_{\mu}$
oscillations is much longer than the diameter of the earth for
typical atmospheric neutrino energies\cite{tale}.
If this is the case then the simple scheme discussed in this
paper may easily be invoked to give
approximately maximal $\nu_\mu \to \nu_\tau$ oscillation solution
for the atmospheric neutrino anomaly in such models.

To be more concrete, let's consider the Exact Parity Model (EPM) 
extended to include a 
heavy see-saw gauge singlet $\nu_R$ (together 
with it's parity partner $\nu '_L$)\cite{flv2}. 
In each generation there are 4 Weyl neutrino fields,
namely, $\nu_L$, $\nu_R$ and their parity partners $\nu '_R$
and $\nu '_L$. With the minimal Higgs and it's mirror 
doublet, the mass matrix has the form\cite{flv2} 
\begin{equation}
{\cal L}_{mass} = \bar \Psi_L {\cal M} (\Psi_L)^c + H.c.,
\label{24} \end{equation}
where 
\begin{equation}
\Psi_L = [(\nu_L)^c,  \nu '_R,  \nu_R, (\nu '_L)^c]^T, \label{25}
\end{equation}
and
\begin{equation}
{\cal M} = \left(\begin{array}{cccc}
0&0&m_a&m_b\\
0&0&m_b&m_a\\
m_a&m_b&M_a&M_b\\
m_a&m_b&M_b&M_a
\end{array}\right). 
\label{26}
\end{equation}
Note that $m_{a,b}$ are mass terms that arise from
spontaneous symmetry breaking, while $M_{a,b}$ are bare mass terms,
all being free parameters of the theory. 
The masses are assumed to be real without loss 
of generality. In the parity diagonal basis 
$\nu^{\pm}_{L} = {\nu_L \pm (\nu '_R)^c \over
\sqrt{2}}$ and $\nu^{\pm}_{R} = {\nu_R \pm (\nu '_L)^c \over
\sqrt{2}}$, the mass matrix of Eq. (\ref{26}) is diagonalised 
to give eigenvalues $m_+, m_-, M_+, M_- $, where $m_\pm, M_\pm $
are functions of $m_{a,b}$ and $M_{a,b}$\cite{flv2}. Following the usual 
nomenclature, $m_+, m_-$ refer to the masses of light neutrino 
parity  states of $\nu^{\pm}_{L}$, whereas $ M_+, M_- $ are the 
masses of the heavy parity states $\nu^{\pm}_{R}$. In the 
limit $ M_ \pm \gg m_ \pm $
the singlet states decouple from the SU(2) 
doublet states. It is straight forward to generalise the 1 generation 
case to describe cross generational mixing among the 2-3 sector.
Introducing 2 arbitrary intergenerational
mixing angles $\theta$ and $\phi$ to parametrise the 
generation mixing between the +(-) parity states $\nu^+_2$ and $\nu^+_3$ 
($\nu^-_2$ and $\nu^-_3$), the active flavour states will have the form
\begin{eqnarray}
\nu_{\mu L} &={\cos\theta \nu_{2L}^+
  \over \sqrt 2} + {\sin\theta \nu_{3L}^+ \over \sqrt 2}
+ {\cos \phi \nu_{2L}^{-} \over \sqrt 2} + {\sin \phi \nu_{3L}^{-}
\over \sqrt 2}\nonumber \\
\nu_{\tau L} &= {-\sin\theta \nu_{2L}^+  \over \sqrt 2} +
{\cos\theta \nu_{3L}^+ \over \sqrt 2}
- {\sin \phi \nu_{2L}^{-} \over \sqrt 2} +
{\cos \phi \nu_{3L}^{-} \over \sqrt 2}
\label{zz}
\end{eqnarray}
From Eq. (\ref{zz}), the transition probability 
from $\nu_{\mu L}$ to $\nu_{\tau L}$ in the relativistic limit
is
\begin{eqnarray}
P(\nu_{\mu L} \rightarrow \nu_{\tau L})
&=& | \langle {\nu_{\tau L} | \nu_{\mu L} (t)} \rangle |^2
\nonumber \\
&=& \frac{1}{4} \sin^2{2\theta} \, \sin^2({\pi L \over L_+}) 
\,+ {1 \over 4} \sin^2{2\phi} \, \sin^2( {\pi L \over L_-})  \nonumber
\\
&+& \frac{1}{2} \sin\,{2\theta}\sin \,{2\phi}
\sin( {\pi L \over L_-} ) 
\sin( {\pi L \over L_+} )\cos \, (\frac{\pi L}{L_{int}}), \label{28}
\end{eqnarray}
where the oscillation lengths are
\begin{equation}
L_\pm  =   {4\pi E \over \delta {m_{32}^\pm}^2}, \
L_{int}  = {4\pi E \over \delta m^2_g}.
\end{equation}
In the above equations, $E$ denotes the energy of 
$\nu_{\mu L}$, $L$ is 
the distance from where it is produced,  
${\delta {m_{32}^\pm}^2} \equiv (m_3^\pm)^2 - (m_2^\pm)^2$ and
$\delta m^2_g \equiv (m_3^+)^2+(m_2^+)^2-(m_2^-)^2-(m_3^-)^2$.
Note that $L_{int}$ and $L_\pm$ are independent of each other
because the mass square scale difference 
$\delta {m_{32}^\pm}^2$ and $\delta m_g^2$ are independent
of each other in the theory.

The limit of the (accidental?) 
$U(1)_{ L_\mu - L_\tau}$ global symmetry would mean 
(i) $\phi \approx \theta \approx \frac {\pi}{4}$ 
and (ii) the eigen masses of 
the two generations will be approximately degenerate, 
i.e.  $(m_3^+)^2 \approx (m_2^+)^2, (m_3^-)^2 \approx (m_2^-)^2$. 
To produce $\nu_\mu - \nu_\tau$ maximal oscillations
in the EPM, we also need to render approximate equality in the oscillation
lengths, $L_+ \approx L_-$ and $L_{int} \gg  L$. 
This happens when 
the coupling between the ordinary and mirror neutrinos is small
(so that $m_3^+ \to m_3^-,\ m_2^+ \to m_2^-$).
In this limit, the standard two-neutrino
oscillation formula 
\begin{equation}
P(\nu_{\mu L} \rightarrow \nu_{\tau L}) 
= \sin^2 2\theta \sin^2 {(\frac{\pi L}{L_+})}
\label{222}
\end{equation}
is then recovered.
Thus for atmospheric neutrinos, which have
typical energy $E \sim \hbox{GeV}$, we require
$L_{int} \gg 10,000 \, \hbox{km}$.
In this limit, the atmospheric $\nu_\mu$ oscillations 
approximately reduce to pure $\nu_\mu \to \nu_\tau$ oscillations.

Thus we have shown that this ``lop sided" ${\bf M}_R$ scenario does fit in
nicely with the EPM. Hence, if experiments do end up proving that
the atmospheric neutrino anomaly is due to $\nu_\mu \to \nu_\tau$
oscillations, and the solar problem is due to maximal 
$\nu_e \to \nu_{sterile}$ oscillations then this is consistent
with the EPM.
Of course, within the framework of the EPM we would most
naturally expect the atmospheric neutrino anomaly to be
solved by maximal $\nu_\mu \to \nu'_\mu$ oscillations, since then
we can solve the atmospheric neutrino anomaly and solar
neutrino problem by the same mechanism.
This would seem to be theoretically most satisfying.

The implications for the early Universe of such models with
sterile/mirror neutrinos have been 
discussed in some detail in the literature (see e.g.
Ref.\cite{fvap} and references there-in).
Models with sterile neutrinos have the amusing feature
that they typically lead to the dynamical generation of
a large neutrino asymmetry\cite{ftv,fv}.
This leads to a number of interesting effects for big bang
nucleosynthesis (BBN), and the microwave background.
For example, it was shown in Ref.\cite{fv} that
the maximal $\nu_e \to \nu'_e$ oscillations do not
necessarily lead to adverse implications for BBN 
even if $\delta m^2$ is relatively large $\sim
10^{-3}\ eV^2$ because of the suppression of the oscillations
due to the dynamically generated $L_{\nu_\mu}, L_{\nu_\tau}$ 
asymmetry.
Further implications for models with degenerate
$\nu_\mu, \nu_\tau$ are discussed in Ref.\cite{bfv}.

In conclusion, we have suggested a simple scheme to
achieve approximately maximal $\nu_\mu - \nu_\tau$ oscillations
which is one of the possible solutions to the atmospheric
neutrino anomaly. In the context of this scheme, we have shown
that the Exact Parity model
could also accommodate $\nu_\mu - \nu_\tau$ oscillations in place 
of the usual $\nu_\mu - \nu '_\mu$ oscillations. 
This scheme is also compatible with the LSND oscillation signal, which leaves
the solar neutrino problem to be solved by $\nu_e \to \nu_{sterile}$
oscillations. 

\vskip 0.5cm
\noindent
{\bf Acknowledgements}
\vskip 0.5cm
R.F. is an Australian Research Fellow.
T. L. Yoon is supported by OPRS and MRS

\end{document}